\documentclass[]{spie}  

\usepackage{amsmath,amsfonts,amssymb}
\usepackage{graphicx}
\usepackage[colorlinks=true, allcolors=blue]{hyperref}
\usepackage{wrapfig}
\usepackage{floatrow}
\usepackage{float}

\title{Distributed Aperture Telescopes and the Dragonfly Telephoto Array}

\author[a,b]{Roberto G. Abraham}
\author[c]{Pieter G. van Dokkum}
\author[d]{Deborah M. Lokhorst}
\author[a,b]{Seery Chen}
\author[a]{Qing Liu}
\author[e]{Michael L. Rice}
\author[e]{E. Lynn Rice}

\affil[a]{David A. Dunlap Department of Astronomy \& Astrophysics,
University of Toronto,
50 St. George Street, 
Toronto, ON M5S3H4, Canada}
\affil[b]{Dunlap Institute for Astronomy \& Astrophysics,
University of Toronto,
50 St. George Street, 
Toronto, ON M5S3H4, Canada}
\affil[c]{Department of Astronomy,
Yale University,
52 Hillhouse Ave., New Haven, CT 06511, USA}
\affil[d]{NRC Herzberg Astronomy \& Astrophysics Research Centre,
5071 West Saanich Road, 
Victoria, BC V9E2E7, Canada}
\affil[e]{New Mexico Skies, Inc., 
9 Contentment Crest, Mayhill, NM 88339, USA}

\authorinfo{Send correspondence to R.G.A. at roberto.abraham@utoronto.ca}

\begin{document} 
\maketitle

\begin{abstract}
Telescope arrays allow high-performance wide-field imaging systems to be built more quickly and at lower cost than conventional telescopes. Distributed aperture telescopes (the premier example of which is the Dragonfly Telephoto Array) are a special type of array in which all telescopes point at roughly the same position in the sky. In this configuration the array performs like a large and optically very fast single telescope with unusually good control over systematic errors. In a few key areas, such as low surface brightness imaging over wide fields of view, distributed aperture telescopes outperform conventional survey telescopes by a wide margin. In these Proceedings we outline the rationale for distributed aperture telescopes, and highlight the strengths and weaknesses of the concept. Areas of observational parameter space in which the design excels are identified. These correspond to areas of astrophysics that are both relatively unexplored and which have unusually strong breakthrough potential.
\end{abstract}


\section{Introduction}

Telescope designs are complicated trade-offs between aperture, field of view, resolution, wavelength coverage, cost, complexity, and many other factors. Appropriately chosen figures of merit can be useful tools for navigating this trade space. For example, in the case of a seeing-limited telescope designed to survey compact sources over a wide area, a suitable figure of merit\index{figure of merit} $\Phi$ would be:
\begin{equation} 
\Phi \propto {{D^2 \Omega\, \eta}\over {d\Omega}},
\label{eqn:meritLSST}
\end{equation}
where $D$ is the aperture, $\Omega$ is the field of view, $\eta$ is the throughput and $d\Omega$ is the resolution. Maximizing this figure of merit drives one toward a telescope design with a fairly large aperture and a large field of view, such as the innovative optical design chosen for the Vera Rubin Telescope\cite{angelDesign8mTelescope2000, seppalaImprovedOpticalDesign2002}. On the other hand,  if one is more interested in investigating individual targets over a small area of the sky at resolutions approaching the diffraction limit ({\em e.g.} using adaptive optics), a more appropriate figure of merit is: 
\begin{equation}
\Phi \propto {{D^4 \Omega\, \eta} \over \lambda},
\label{eqn:meritELT}
\end{equation}
\noindent where $\lambda$ is the wavelength. Maximizing the aperture to exploit the $D^4$ factor in this expression\cite{nelsonStatusThirtyMeter2008,hookReportESOWorkshop2009} lies at the heart of the designs  for the European Extremely Large Telescope and the Thirty Meter Telescope. 

Expressions (\ref{eqn:meritLSST}) and (\ref{eqn:meritELT}) bookend most science cases. However, there are interesting situations where neither figure of merit is particularly appropriate. For example, consider the case where one is only interested in imaging very faint structures that are much larger than the limiting resolution of the instrument (or the atmosphere).  In this case it is arguable that the appropriate figure of merit scales with the number of detected photons per pixel from the extended source:
\begin{equation}
\Phi \propto \mu\, a\, \eta f^{-2},
\label{eqn:merit}
\end{equation}
where $f$ is the focal ratio, $\mu$ is the surface brightness and $a$ is the pixel area. This relationship is interesting because the aperture of the telescope does not enter into the expression directly, but the focal ratio does\cite{abrahamUltraLowSurface2014}. Therefore, for imaging very extended low surface brightness structures ({\em i.e.} ones far larger than the resolution limit), an {\em optically fast} (low $f$-ratio) telescope may be preferable to a large aperture telescope. Since telescope arrays reduce the effective focal ratio of the imaging system (more on this below), this line of thinking suggests that {\em in certain situations} arrays of small telescopes may outperform monolithic aperture telescopes. 

Telescope arrays can be used in two modes\footnote{Of course in these Proceedings we are referring to arrays operating at visible wavelengths. Radio astronomers have been using telescope arrays for over half a century, operating primarily in other modes, such as interferometrically.}. In the first, the elements of the array are adjusted so that their areal footprints do not overlap. The main utility of an array operating in this mode is to enlarge the instrument's instantaneous field of view. This is the mode in which arrays designed to identify optical transients (such as SuperWASP\cite{pollaccoWASPProjectSuperWASP2006}, TESS\cite{rickerTransitingExoplanetSurvey2014,chrispOpticalDesignCamera2015}, EvryScope\cite{lawEvryscopeScienceExploring2015}, ASAS-SN\cite{kochanekAllSkyAutomatedSurvey2017}, and many others) operate. The trade-offs inherent to such systems are well-understood, and we refer the reader to Ackermann et al. 2016 for a comprehensive summary\cite{ackermannLensCameraArrays2016}. The second mode, in which the areal footprints are adjusted so they mostly (but not perfectly) overlap, is less well-explored. We will refer to this as the `distributed aperture telescope' concept, and sketching out the strengths and weaknesses of arrays operating as distributed aperture telescopes is the focus of the present article. 

\noindent Note: this paper is one of three in these proceedings describing the development of the Dragonfly Telephoto Array. Table \ref{tab:3papers} summarizes the topics covered by each paper. 

\begin{table}[ht]
\caption{Content summary of the three papers in this series in these Proceedings. Topics covered by the present paper are highlighted in bold.} 
\label{tab:3papers}
\begin{center}       
\begin{tabular}{|l|l|} 

\hline
\rule[-1ex]{0pt}{3.5ex}  \textbf{Distributed aperture telescope general concepts} & This article \\
\hline
\rule[-1ex]{0pt}{3.5ex}  \textbf{Low surface brightness imaging challenges}	 & This article \\
\hline
\rule[-1ex]{0pt}{3.5ex}  \textbf{Lessons learned from Dragonfly} & This article  \\
\hline
\rule[-1ex]{0pt}{3.5ex}  Narrowband imaging concepts and methods & Lokhorst et al.  \\
\hline
\rule[-1ex]{0pt}{3.5ex}  Narrowband imaging survey speed & Lokhorst et al. \\
\hline
\rule[-1ex]{0pt}{3.5ex}  Dragonfly Spectral Line Mapper pathfinder results \& lessons learned & Lokhorst et al. \\
\hline
\rule[-1ex]{0pt}{3.5ex}  Dragonfly Spectral Line Mapper design  & Chen et al. \\
\hline
\rule[-1ex]{0pt}{3.5ex}  Dragonfly Spectral Line Mapper laboratory tests	 & Chen et al.\\
\hline
\rule[-1ex]{0pt}{3.5ex}  Dragonfly Spectral Line Mapper roadmap & Chen et al. \\
\hline
\end{tabular}
\end{center}
\end{table} 

\section{Distributed Aperture Telescope Arrays vs. Mosaic Sensor Arrays}

Assuming interferometric methods are not being used, and that sub-exposure stacking is used to construct the final image, a distributed aperture telescope array can never provide resolution better than that of a single element in the array. However, an eight-inch telescope ($D\sim20~\rm{cm}$) operating in V-band already has a diffraction-limited angular resolution of $\lambda/D\sim0.6~\rm{arcsec}$, which corresponds to good natural seeing on all of the world's best sites. This makes it clear that in vast majority of cases, at visible wavelengths, the main benefit of large aperture telescopes is that they collect more photons, not that they improve angular resolution. In such cases, there is no {\em fundamental} difference between obtaining an image with a large aperture telescope and stacking images (obtained at the same time) from an array of smaller telescopes. Whether or not there is a {\em practical} difference depends on a myriad number of factors, such as the read noise and dark current in the detectors relative to the poisson noise from the sky background, and more will be said about these considerations below. However, at least in principle\cite{abrahamUltraLowSurface2014}, a stacked image from a telescope array is equivalent to that obtained from a ground-based telescope with aperture $D_{\rm eff}$ and focal ratio $f_{\rm eff}$:
\begin{eqnarray}
D_{\rm eff} & =\sqrt{N} \times D \label{eqn:arrayD}\\
f_{\rm eff} & = f/\sqrt{N} \label{eqn:arrayF},
\end{eqnarray}

\noindent where $N$ is the number of telescopes in the array, each of which has aperture $D$ and focal ratio $f$. In general, the lenses in the array are only approximately co-aligned (with small offsets corresponding to $5-10\%$ of the field of view). This ensures that the optical path of the wavefront through each element in the array is slightly different, which greatly aids in the removal of scattered light and internal reflections in the final stacked image. Therefore the instantaneous field of view of the array is similar to the instantaneous field of view of an individual element in the array.  

The decrease in the effective focal ratio of the array show by Equation~\ref{eqn:arrayF} is interesting, since Expression~(\ref{eqn:merit}) suggests that focal ratio matters more than aperture for imaging very large structures down to low surface brightness levels. This line of argument clearly has some truth to it, but it can also be taken too far. The main technological factor making telescope arrays a practical reality is the fact that images from digital sensors can be stacked easily. Since digital sensors can also be re-binned easily, it is arguable that the best way to optimize this figure of merit is to simply bin the sensor (increasing pixel area $a$). Of course, this is only an option if the target fits within the field of view! For this reason, we think that et\'endue conservation ultimately provides a better justification for distributed aperture telescopes, based on the following argument. 

\begin{figure}[htb]
\includegraphics[width=16cm]{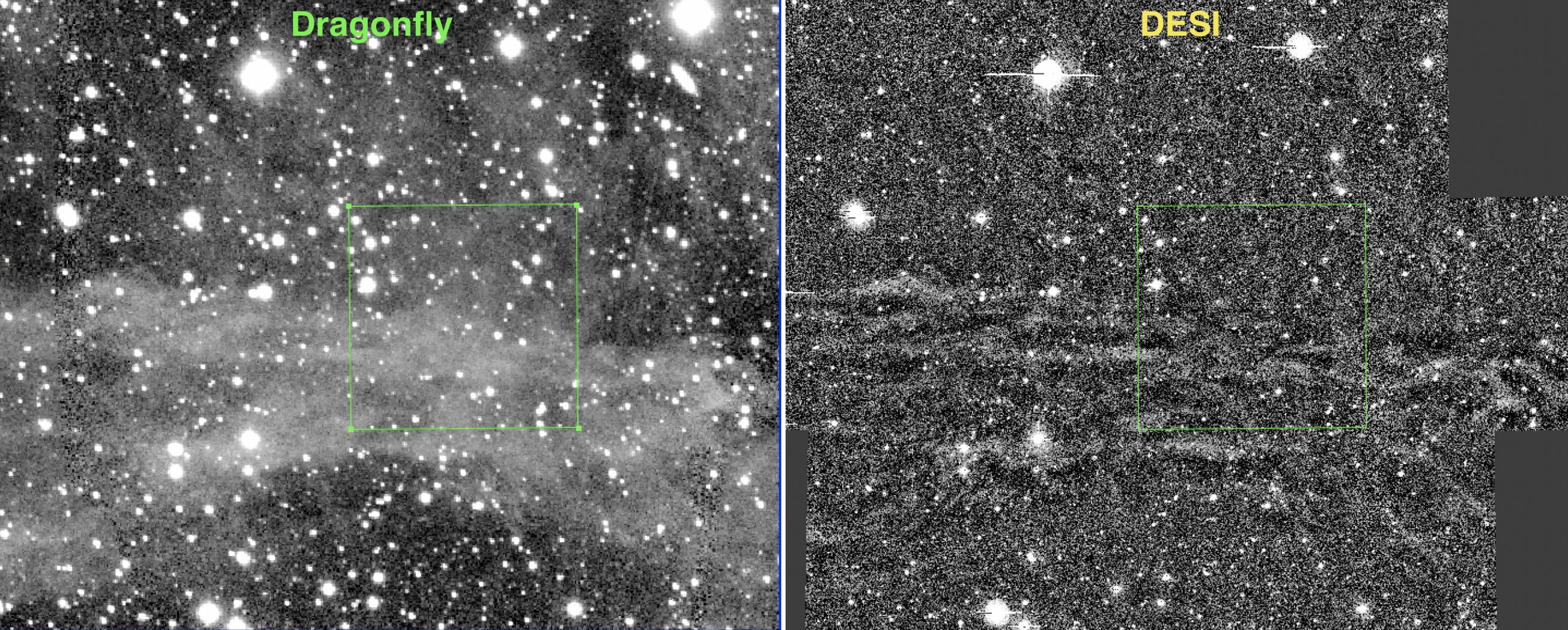}
\caption{A 45x50 arcmin$^2$ central region of SPIDER field, a diffuse HI gas region at high Galactic latitude, imaged by Dragonfly (left) and by the DESI Legacy Imaging survey (right). Exposure times are similar. In the DESI image, low frequency power is suppressed by the sky modelling algorithm needed to construct the image from the sensor mosaic. Individual sensors have a footprint of 0.25° × 0.25°, as indicated by the the green squares in both panels, and structures on this scale are removed. Dragonfly's distributed aperture design is in some sense a `mosaic {\em telescope} array' (not a mosaic {\em sensor} array), so individual sensors cover very large fields of view (several square degrees), and low surface brightness structures are well preserved. }
\label{fig:desi}
\end{figure}

The angular field of view $\omega$ of an imaging system with focal length $l$ and sensor size $H$ is given by:
\begin{equation}
\omega = 2 \tan^{-1}\left({H\over 2 l}\right).
\label{eqn:fov}
\end{equation}
Et\'endue conservation limits the focal ratio of any (lossless) optical system to $f \ge 0.5$, so for an {\em optimal} wide-field imager (i.e. one that is as optically fast as conservation of energy allows), $l = D/2$. Assuming such a perfectly fast optical  design, (\ref{eqn:fov}) can be rewritten as:
\begin{equation}
\omega_{\rm opt} = 2 \tan^{-1}\left({H\over D}\right)
\label{eqn:optfov}
\end{equation}
where $\omega_{\rm opt}$ is the maximum field of view achievable with a telescope of aperture $D$ and sensor size $H$. This equation shows that to obtain a wide field of view with a large telescope, every increase in aperture {\em must} be accompanied by a commensurate increase in the size of a sensor.  In the best possible case, the constant of proportionality is unity, and doubling the aperture means doubling the size of the sensor if the field of view is to be preserved. In more realistic cases, the sensor grows as some multiple of the aperture increase. No clever optical design can ever sidestep the fact that large telescopes with wide fields of view must also have huge sensors. This is a major problem, because the technological trend for sensor development over the last couple of decades has been to move in the opposite direction, namely toward {\em miniaturization} of sensors. At the time of writing, a few dollars can easily buy a small off-the-shelf 20\,megapixel sensor that would provide a very wide field of view on a small telescope. These off-the-shelf sensors have quantum efficiencies comparable to that of the best monolithic CCD sensors for 8m-class telescope, but at a cost that is about four orders of magnitude lower, because manufacturing of large CCD sensors has been nearly stagnant for almost two decades. 

The community's cure for the problem of requiring larger and larger sensors as aperture increases is to use mosaiced sensor arrays at the focal planes of very complicated re-imaging cameras fed by large telescopes. This approach takes advantage of our existing arsenal of large telescopes, but it is not without drawbacks. For example, high precision photometry is difficult unless individual targets fit within the areal footprint of an individual sensor, because sky-subtraction operates on the scale of individual devices. Images from mosaiced CCD cameras are thus effectively high-pass filtered (Figure~\ref{fig:desi}).\footnote{The degree to which chip-by-chip background modelling is absolutely necessary is debatable, but in any case, at present all mosaic sensor pipelines rely on it.} Our view is that for science cases where such filtering matters, and where high-precision wide-area low surface brightness photometry is important, a more robust approach is to use a distributed aperture telescope array. In fact, the ideal use case is to use both types of telescope in complementary ways\cite{vandokkumMultiresolutionFilteringEmpirical2020,martinPreparingLowSurface2022a}.

\begin{figure}
\floatbox[{\capbeside\thisfloatsetup{capbesideposition={right,top},capbesidewidth=7cm}}]{figure}[\FBwidth]
{\caption{One of the two 24-lens arrays comprising the 48-lens Dragonfly Telephoto Array. The lenses are co-aligned and the full array is equivalent to a 1m aperture $f/0.39$ refractor with a $2\times 3$\,deg field of view. The arrays are housed in domes at the New Mexico Skies observatory, and operate as a single telescope slaved to the same robotic control system. The lenses are commercial 400\,mm Canon USM IS~II lenses that have superb (essentially diffraction limited) optical quality. This particular lens has very low scatter on account of proprietary nanostructure coatings on key optical surfaces (see Abraham \& van Dokkum 2014 for details). Each lens is affixed to a separate CCD camera and both are controlled by a miniature computer attached to the back of each camera that runs bespoke camera and lens control software that we have made publicly available on a GitHub repository. In the latest incarnation of Dragonfly, each lens is self-configuring and is controlled by its own node.js JavaScript server  in an ``Internet of Things'' configuration that provides a RESTful interface. Growing the array is done by simply bolting a new lens onto the array and plugging in network and power cables.}\label{fig:nms}}
{\includegraphics[width=8cm]{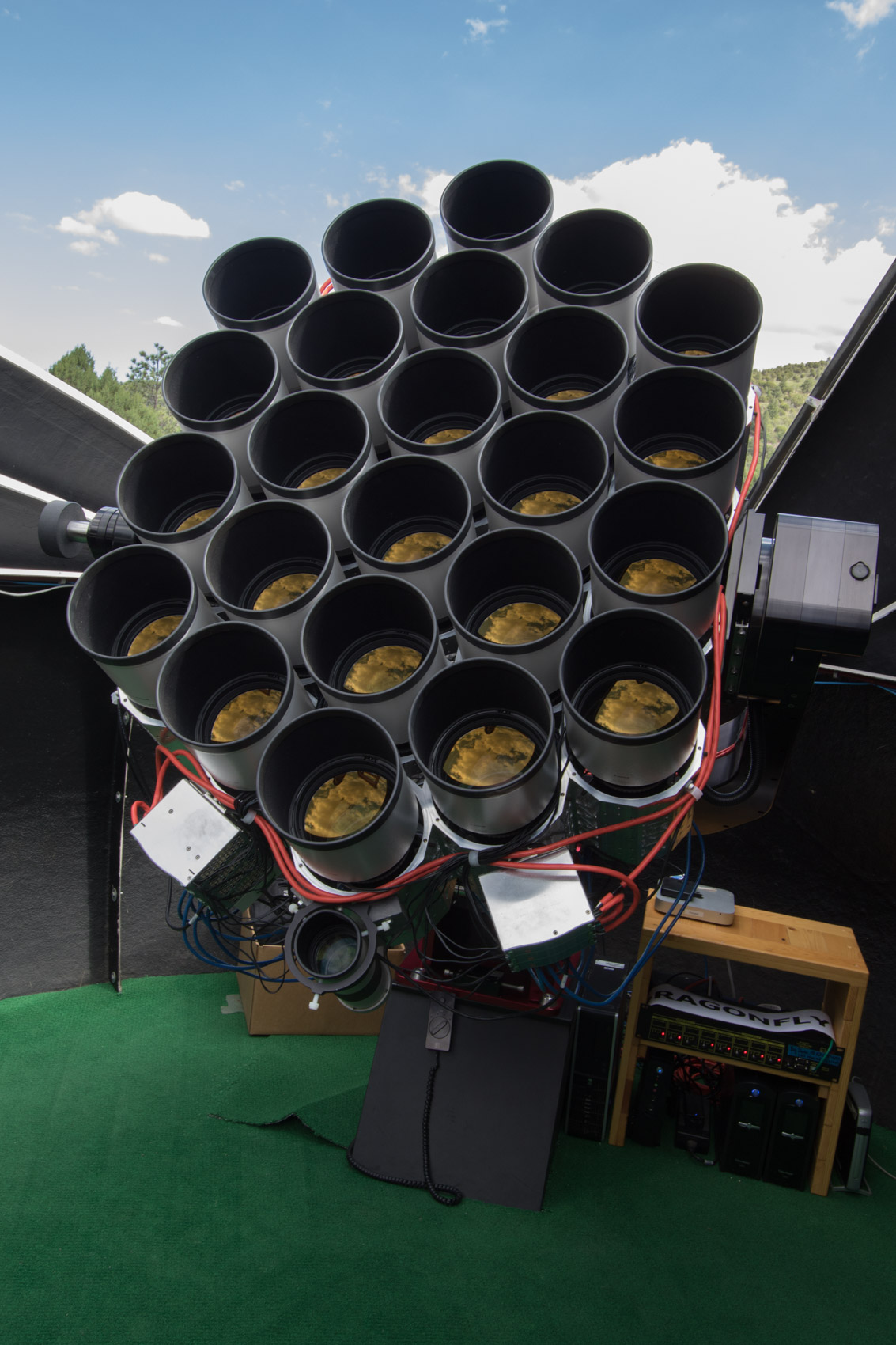}}
\end{figure}

\section{The Dragonfly Telephoto Array}

The Dragonfly Telephoto Array\cite{abrahamUltraLowSurface2014} (Dragonfly for short) is currently the premier example of a telescope built with the ideas expressed in the previous section in mind. An added element in the design of Dragonfly is that it is optimized to explore the low surface brightness Universe with high precision. The most intractable problem in low surface brightness imaging is scattering in the optical train, typically from faint stars in the field, though sometimes from bright stars outside the field\cite{slaterRemovingInternalReflections2009a,euclidcollaborationEuclidPreparationXVI2022}. Because this scattering originates in several of the basic design trades that make large telescopes possible (e.g., an obstructed pupil and reflective surfaces that have high-frequency micro-roughness that backscatters into the optical path and pollutes the focal plane\cite{nelsonAnalysisScatteredLight2007}), Dragonfly has an unobstructed pupil and no reflective surfaces at all. The array builds up its effective aperture by multiplexing the latest generation of high-end commercial telephoto lenses that use nano-fabricated coatings with sub-wavelength structures to yield a factor of ten improvement in wide-angle scattered light relative to other astronomical telescopes\cite{liuMethodCharacterizeWideangle2022,sandinInfluenceDiffuseScattered2015}.  The array is designed to increase in aperture with time, and over the last eight years a 48-lens Dragonfly array has been assembled gradually in New Mexico as a collaboration between the University of Toronto, Yale and Harvard. In its current configuration (half of which is shown in Figure~\ref{fig:nms}) Dragonfly is  equivalent to a 1\,m aperture $f/0.39$ refracting telescope with a six square degree field of view and optical scattering an order of magnitude lower than conventional telescopes. The current detectors have $5.4\,\mu$m pixels resulting in an angular resolution of 2.85\,arcsec/pixel, so the images are under sampled but the array has superb performance when imaging low surface brightness structures on scales larger than about 8\,arcsec. 

Dragonfly is the world’s most powerful wide-field low surface brightness imager, and routinely probes diffuse structures to surface brightness limits well below 30 mag/arcsec$^2$. Within months of commissioning, the array discovered a surprising diversity in stellar halos around nearby massive galaxies\cite{dokkumFirstResultsDragonfly2014,gilhulyStellarHalosDragonfly2022a} and an abundance of ultra-diffuse galaxies (a term coined by our team) in the Coma cluster\cite{vandokkumFORTYSEVENMILKYWAYSIZED2015}, leading to numerous other investigations of ultra-diffuse systems\cite{kodaApproximatelyThousandUltradiffuse2015,janssensUltradiffuseUltracompactGalaxies2017a,iodiceFirstDetectionUltradiffuse2020,marleauUltraDiffuseGalaxies2021,forbesUltradiffuseGalaxiesIC2020,barbosaOneHundredSMUDGes2020,zaritskySystematicallyMeasuringUltraDiffuse2022}. Dragonfly discoveries have been followed up with the world’s largest telescopes to reveal remarkable properties of galaxies, including systems with anomalously high dark matter abundances\cite{vandokkumSpatiallyResolvedStellar2019}, and galaxies with no dark matter at all\cite{vandokkumGalaxyLackingDark2018,danieliStillMissingDark2019}. Dragonfly is now undertaking a wide-area ultra-deep survey of the local volume, and (as described by Lokhorst et al. and Chen et al. in these Proceedings) is simultaneously being upgraded to 168 lenses, 120 of which will have ultra-narrow-band imaging filters designed to allow direct imaging of the circum-galactic medium of galaxies, the ‘hidden’ reservoir of most of the Universe’s baryons\cite{lokhorstWidefieldUltranarrowbandpassImaging2020,lokhorstGiantShellIonized2022}. 

\section{Some lessons from Dragonfly}

The discussion in Sections~1 and 2 presents the advantages and disadvantages of distributed aperture telescopes in a rather abstract way. We have now been operating Dragonfly for a number of years, so it seems timely to augment these abstractions with a few lessons learned from `the school of hard knocks' that might prove useful to others contemplating the construction of a similar system.  
 
The primary disadvantage of distributed aperture telescopes is clearly their operational complexity.  Operating $N$ telescopes with $N$ instruments means there are roughly $2N$ more things that can go wrong.  The best defence against such complexity is parallelism, because if $N$ is large and all telescopes are completely independent, then it does not much matter if a few telescopes are non-operational on a given night. Another potential disadvantage is the relatively complicated reduction pipeline required to handle the large data volumes produced. However, the last few decades have seen innovations in computation far out-stripping innovations in classical optics, so relying on Moore's law and the inherent parallelism of image stacking means that computational complexity is no longer a serious disadvantage, at least not when compared to mechanical complexity.
 
The complexity of distributed aperture telescope arrays is somewhat offset by the optical simplicity of individual elements. Individual telescopes are small enough that simple designs offer good performance. In fact, all-refractive designs (such as used by Dragonfly) are feasible for programs in which scattered light and ghosting are the limiting systemics. And by making $N$ large enough, systems with $f_{\rm eff} < 0.5$ are possible, which is simply not possible using monolithic telescopes (since, as noted earlier, a focal ratio of $f=0.5$ is a hard thermodynamic limit imposed by et\'endue conservation). The relatively small entrance pupils of individual elements of the array also means that full-aperture narrow-bandpass interference filters can be used in some cases, with tremendous improvements in performance (see articles by Deborah Lokhorst et al. and Seery Chen et al. in these proceedings for detailed examples).

Mechanical flexure in the array is emerging as an important limiting factor for Dragonfly. Distributed aperture telescopes require stable optical axes for all elements of the array during exposures. The existing Dragonfly design is adequately rigid over the relatively short (300s to 600s) integrations needed to obtain sky-limited integrations with broad-band filters. However, as Dragonfly evolves to encompass ultra-narrow-band imaging capability, the required integration times become very long, and internal flexure results in distorted images. Our planned solution for this is to eventually replace Dragonfly's CCD cameras with low read-noise CMOS cameras. This will greatly decrease the sub-exposure integration time needed to be sky-noise limited. However, after evaluating a number of current-generation commercial CMOS cameras, we have come to the conclusion that they do not yet have the electrical stability of the best-of-breed commercial CCD cameras, and the additional electronics that they require results in significant amplifier glow on images.  We are therefore giving CMOS cameras more time to mature before moving to this technology. In the meantime, flexure is being addressed by replacing existing cameras with lower read-noise CCD cameras, and by adding a closed-loop tip-tilt active control system to each element of the array. 

Another lesson from Dragonfly is that some of the sources of systematic error in ultra-deep imaging find their origins in the atmosphere, rather than in the telescope. To combat these, the operational model for using Dragonfly is in some ways as innovative as the hardware. When investigating galaxy haloes, the array points only at locations pre-determined (on the basis of IRAS 100$\mu$ imaging) to have low Galactic cirrus contamination, and the array operates in a fully autonomous robotic mode that tracks atmospheric systematics in real time. The latter point is important because once one has greatly reduced the wide-angle scatter inherent to the instrument, the tall pole becomes the atmosphere. The wide-angle telescopic PSF (the ``stellar auroeole'') is not well understood and at least some of its origin is instrumental\cite{racineTelescopicPointSpreadFunction1996,bernsteinOpticalExtragalacticBackground2007}. But a very significant fraction is due to scattering by icy aerosols in the upper atmosphere (a fact well known to atmospheric physicists, and with some implications for climate change models\cite{devoreRetrievingCirrusMicrophysical2013}). This component of the wide-angle scatter is variable\cite{sandinInfluenceDiffuseScattered2014}, and our Dragonfly data shows quite clearly that this variability extends down to a timescale of minutes. It is possible to account for wide-angle PSF contamination via post-facto measurement and subsequent modelling\cite{slaterRemovingInternalReflections2009a,trujillo31MagArcsec2016a,gilhulyStellarHalosDragonfly2022a} and we have developed software tools for this purpose\cite{liuMethodCharacterizeWideangle2022}. Until the properties of the wide-angle PSF are better understood, real-time monitoring of the large-angle PSF is likely to be important for reliable measurement of low surface brightness features on angular scales of arcminutes and larger. 

The final very significant lesson from Dragonfly is that distributed aperture telescope arrays can start small and grow large quickly and relatively inexpensively. For telescope arrays, the {\em worst-case} cost scaling with aperture is $D_{\rm eff}^2$ for both the telescope {\em and} the enclosure, and real-world cost scaling is much more favourable because of reliance on mass produced components which benefit from economies of scale. Dragonfly started off with three lenses on a single mount, is presently 48 lenses spread over two mounts, and is scheduled to grow to 168 lenses on six mounts by end of 2023. 

\acknowledgments 
 
Construction and operation of the Dragonfly Telephoto Array is made possible by support from the Dunlap Institute (funded through an endowment established by the David Dunlap family and the University of Toronto), the Natural Sciences and Engineering Research Council of Canada, the U.S. National Science Foundation, the National Research Council of Canada, and the Canada Foundation for Innovation. We are very grateful to the staff at New Mexico Skies, Inc. for helping to build and operate Dragonfly.
\bibliography{references} 

\begin{thebibliography}{10}

\bibitem{angelDesign8mTelescope2000}
Angel, R., Lesser, M., Sarlot, R., and Dunham, E., ``Design for an 8-m
  {Telescope} with a 3 {Degree} {Field} at f/1.25: {The} {Dark} {Matter}
  {Telescope},'' ~{\bf 195},  81 (Jan. 2000).
\newblock Conference Name: Imaging the Universe in Three Dimensions ADS
  Bibcode: 2000ASPC..195...81A.

\bibitem{seppalaImprovedOpticalDesign2002}
Seppala, L.~G., ``Improved optical design for the {Large} {Synoptic} {Survey}
  {Telescope} ({LSST}),''  111 (Dec. 2002).

\bibitem{nelsonStatusThirtyMeter2008}
Nelson, J. and Sanders, G.~H., ``The status of the {Thirty} {Meter} {Telescope}
  project,''  70121A (Aug. 2008).

\bibitem{hookReportESOWorkshop2009}
Hook, I., Liske, J., Villegas, D., and Kissler-Patig, M., ``Report on the {ESO}
  {Workshop} {E}-{ELT} {Design} {Reference} {Mission} and {Science} {Plan},''
  {\em The Messenger}~{\bf 137},  51--52 (Sept. 2009).
\newblock ADS Bibcode: 2009Msngr.137...51H.

\bibitem{abrahamUltraLowSurface2014}
Abraham, R.~G. and van Dokkum, P.~G., ``Ultra–{Low} {Surface} {Brightness}
  {Imaging} with the {Dragonfly} {Telephoto} {Array},'' {\em Publications of
  the Astronomical Society of the Pacific}~{\bf 126},  55--69 (Jan. 2014).

\bibitem{pollaccoWASPProjectSuperWASP2006}
Pollacco, D., Skillen, I., Cameron, A., Christian, D., Hellier, C., Irwin, J.,
  Lister, T., Street, R., West, R., Anderson, D., Clarkson, W., Deeg, H.,
  Enoch, B., Evans, A., Fitzsimmons, A., Haswell, C., Hodgkin, S., Horne, K.,
  Kane, S., Keenan, F., Maxted, P., Norton, A., Osborne, J., Parley, N., Ryans,
  R., Smalley, B., Wheatley, P., and Wilson, D., ``The {WASP} {Project} and the
  {SuperWASP} {Cameras},'' {\em PUBL ASTRON SOC PAC}~{\bf 118},  1407--1418
  (Oct. 2006).

\bibitem{rickerTransitingExoplanetSurvey2014}
Ricker, G.~R., Winn, J.~N., Vanderspek, R., Latham, D.~W., Bakos, G.~A., Bean,
  J.~L., Berta-Thompson, Z.~K., Brown, T.~M., Buchhave, L., Butler, N.~R.,
  Butler, R.~P., Chaplin, W.~J., Charbonneau, D., Christensen-Dalsgaard, J.,
  Clampin, M., Deming, D., Doty, J., De~Lee, N., Dressing, C., Dunham, E.~W.,
  Endl, M., Fressin, F., Ge, J., Henning, T., Holman, M.~J., Howard, A.~W.,
  Ida, S., Jenkins, J.~M., Jernigan, G., Johnson, J.~A., Kaltenegger, L.,
  Kawai, N., Kjeldsen, H., Laughlin, G., Levine, A.~M., Lin, D., Lissauer,
  J.~J., MacQueen, P., Marcy, G., McCullough, P.~R., Morton, T.~D., Narita, N.,
  Paegert, M., Palle, E., Pepe, F., Pepper, J., Quirrenbach, A., Rinehart,
  S.~A., Sasselov, D., Sato, B., Seager, S., Sozzetti, A., Stassun, K.~G.,
  Sullivan, P., Szentgyorgyi, A., Torres, G., Udry, S., and Villasenor, J.,
  ``Transiting {Exoplanet} {Survey} {Satellite},'' {\em J. Astron. Telesc.
  Instrum. Syst}~{\bf 1},  014003 (Oct. 2014).

\bibitem{chrispOpticalDesignCamera2015}
Chrisp, M., Clark, K., Primeau, B., Dalpiaz, M., and Lennon, J., ``Optical
  design of the camera for {Transiting} {Exoplanet} {Survey} {Satellite}
  ({TESS}),''  96020C (Sept. 2015).

\bibitem{lawEvryscopeScienceExploring2015}
Law, N.~M., Fors, O., Ratzloff, J., Wulfken, P., Kavanaugh, D., Sitar, D.~J.,
  Pruett, Z., Birchard, M.~N., Barlow, B.~N., Cannon, K., Cenko, S.~B., Dunlap,
  B., Kraus, A., and Maccarone, T.~J., ``Evryscope {Science}: {Exploring} the
  {Potential} of {All}-{Sky} {Gigapixel}-{Scale} {Telescopes},'' {\em
  Publications of the Astronomical Society of the Pacific}~{\bf 127},  234--249
  (Mar. 2015).

\bibitem{kochanekAllSkyAutomatedSurvey2017}
Kochanek, C.~S., Shappee, B.~J., Stanek, K.~Z., Holoien, T. W.-S., Thompson,
  T.~A., Prieto, J.~L., Dong, S., Shields, J.~V., Will, D., Britt, C.,
  Perzanowski, D., and Pojmański, G., ``The {All}-{Sky} {Automated} {Survey}
  for {Supernovae} ({ASAS}-{SN}) {Light} {Curve} {Server} v1.0,'' {\em
  PASP}~{\bf 129},  104502 (Oct. 2017).

\bibitem{ackermannLensCameraArrays2016}
Ackermann, M.~R., Cox, D.~D., McGraw, J.~T., and Zimmer, P.~C., ``Lens and
  {Camera} {Arrays} for {Sky} {Surveys} and {Space} {Surveillance},'' Sandia
  {Report} SAND2016-8077, Sandia National Laboratories (2016).

\bibitem{vandokkumMultiresolutionFilteringEmpirical2020}
van Dokkum, P., Lokhorst, D., Danieli, S., Li, J., Merritt, A., Abraham, R.,
  Gilhuly, C., Greco, J.~P., and Liu, Q., ``Multi-resolution {Filtering}: {An}
  {Empirical} {Method} for {Isolating} {Faint}, {Extended} {Emission} in
  {Dragonfly} {Data} and {Other} {Low} {Resolution} {Images},'' {\em PASP}~{\bf
  132},  074503 (July 2020).

\bibitem{martinPreparingLowSurface2022a}
Martin, G., Bazkiaei, A.~E., Spavone, M., Iodice, E., Mihos, J.~C., Montes, M.,
  Benavides, J.~A., Brough, S., Carlin, J.~L., Collins, C.~A., Duc, P.~A.,
  Gómez, F.~A., Galaz, G., Hernández-Toledo, H.~M., Jackson, R.~A., Kaviraj,
  S., Knapen, J.~H., Martínez-Lombilla, C., McGee, S., O’Ryan, D., Prole,
  D.~J., Rich, R.~M., Román, J., Shah, E.~A., Starkenburg, T.~K., Watkins,
  A.~E., Zaritsky, D., Pichon, C., Armus, L., Bianconi, M., Buitrago, F.,
  Busá, I., Davis, F., Demarco, R., Desmons, A., García, P., Graham, A.~W.,
  Holwerda, B., Hon, D. S.~H., Khalid, A., Klehammer, J., Klutse, D.~Y., Lazar,
  I., Nair, P., Noakes-Kettel, E.~A., Rutkowski, M., Saha, K., Sahu, N., Sola,
  E., Vázquez-Mata, J.~A., Vera-Casanova, A., and Yoon, I., ``Preparing for
  low surface brightness science with the {Vera} {C}. {Rubin} {Observatory}:
  {Characterization} of tidal features from mock images,'' {\em Monthly Notices
  of the Royal Astronomical Society}~{\bf 513},  1459--1487 (Apr. 2022).

\bibitem{slaterRemovingInternalReflections2009a}
Slater, C., Harding, P., and Mihos, C., ``Removing {Internal} {Reflections}
  from {Deep} {Imaging} {Datasets},'' {\em Publications of the Astronomical
  Society of the Pacific}~{\bf 121}(885),  15 (2009).

\bibitem{euclidcollaborationEuclidPreparationXVI2022}
{Euclid Collaboration}, Borlaff, A.~S., Gómez-Alvarez, P., Altieri, B.,
  Marcum, P.~M., Vavrek, R., Laureijs, R., Kohley, R., Buitrago, F.,
  Cuillandre, J.-C., Duc, P.-A., Gaspar~Venancio, L.~M., Amara, A., Andreon,
  S., Auricchio, N., Azzollini, R., Baccigalupi, C., Balaguera-Antolínez, A.,
  Baldi, M., Bardelli, S., Bender, R., Biviano, A., Bodendorf, C., Bonino, D.,
  Bozzo, E., Branchini, E., Brescia, M., Brinchmann, J., Burigana, C., Cabanac,
  R., Camera, S., Candini, G.~P., Capobianco, V., Cappi, A., Carbone, C.,
  Carretero, J., Carvalho, C.~S., Casas, S., Castander, F.~J., Castellano, M.,
  Castignani, G., Cavuoti, S., Cimatti, A., Cledassou, R., Colodro-Conde, C.,
  Congedo, G., Conselice, C.~J., Conversi, L., Copin, Y., Corcione, L., Coupon,
  J., Courtois, H.~M., Cropper, M., Da~Silva, A., Degaudenzi, H.,
  Di~Ferdinando, D., Douspis, M., Dubath, F., Duncan, C. A.~J., Dupac, X.,
  Dusini, S., Ealet, A., Fabricius, M., Farina, M., Farrens, S., Ferreira,
  P.~G., Ferriol, S., Finelli, F., Flose-Reimberg, P., Fosalba, P., Frailis,
  M., Franceschi, E., Fumana, M., Galeotta, S., Ganga, K., Garilli, B., Gillis,
  B., Giocoli, C., Gozaliasl, G., Graciá-Carpio, J., Grazian, A., Grupp, F.,
  Haugan, S. V.~H., Holmes, W., Hormuth, F., Jahnke, K., Keihanen, E.,
  Kermiche, S., Kiessling, A., Kilbinger, M., Kirkpatrick, C.~C., Kitching, T.,
  Knapen, J.~H., Kubik, B., Kümmel, M., Kunz, M., Kurki-Suonio, H., Liebing,
  P., Ligori, S., Lilje, P.~B., Lindholm, V., Lloro, I., Mainetti, G., Maino,
  D., Mansutti, O., Marggraf, O., Markovic, K., Martinelli, M., Martinet, N.,
  Martínez-Delgado, D., Marulli, F., Massey, R., Maturi, M., Maurogordato, S.,
  Medinaceli, E., Mei, S., Meneghetti, M., Merlin, E., Metcalf, R.~B., Meylan,
  G., Moresco, M., Morgante, G., Moscardini, L., Munari, E., Nakajima, R.,
  Neissner, C., Niemi, S.~M., Nightingale, J.~W., Nucita, A., Padilla, C.,
  Paltani, S., Pasian, F., Patrizii, L., Pedersen, K., Percival, W.~J.,
  Pettorino, V., Pires, S., Poncet, M., Popa, L., Potter, D., Pozzetti, L.,
  Raison, F., Rebolo, R., Renzi, A., Rhodes, J., Riccio, G., Romelli, E.,
  Roncarelli, M., Rosset, C., Rossetti, E., Saglia, R., Sánchez, A.~G.,
  Sapone, D., Sauvage, M., Schneider, P., Scottez, V., Secroun, A., Seidel, G.,
  Serrano, S., Sirignano, C., Sirri, G., Skottfelt, J., Stanco, L., Starck,
  J.~L., Sureau, F., Tallada-Crespí, P., Taylor, A.~N., Tenti, M., Tereno, I.,
  Teyssier, R., Toledo-Moreo, R., Torradeflot, F., Tutusaus, I., Valentijn,
  E.~A., Valenziano, L., Valiviita, J., Vassallo, T., Viel, M., Wang, Y.,
  Weller, J., Whittaker, L., Zacchei, A., Zamorani, G., and Zucca, E.,
  ``\textit{{Euclid}} preparation: {XVI}. {Exploring} the ultra-low surface
  brightness {Universe} with \textit{{Euclid}} /{VIS},'' {\em A\&A}~{\bf 657},
  A92 (Jan. 2022).

\bibitem{nelsonAnalysisScatteredLight2007}
Nelson, P.~G., ``An {Analysis} of {Scattered} {Light} in {Reflecting} and
  {Refracting} {Primary} {Objectives} for {Coronagraphs},'' tech. rep., Coronal
  Solar Magnetism Observatory (2007).

\bibitem{liuMethodCharacterizeWideangle2022}
Liu, Q., Abraham, R., Gilhuly, C., van Dokkum, P., Martin, P.~G., Li, J.,
  Greco, J.~P., Lokhorst, D., Chen, S., Danieli, S., Keim, M.~A., Merritt, A.,
  Miller, T.~B., Pasha, I., Polzin, A., Shen, Z., and Zhang, J., ``A {Method}
  to {Characterize} the {Wide}-angle {Point}-{Spread} {Function} of
  {Astronomical} {Images},'' {\em ApJ}~{\bf 925},  219 (Feb. 2022).

\bibitem{sandinInfluenceDiffuseScattered2015}
Sandin, C., ``The influence of diffuse scattered light. {II}. {Observations} of
  galaxy haloes and thick discs and hosts of blue compact galaxies,'' {\em
  {\textbackslash}aap}~{\bf 577},  A106 (2015).

\bibitem{dokkumFirstResultsDragonfly2014}
Dokkum, P. G.~v., Abraham, R., and Merritt, A., ``First {Results} {From} the
  {Dragonfly} {Telephoto} {Array}: the {Apparent} {Lack} of a {Stellar} {Halo}
  in the {Massive} {Spiral} {Galaxy} {M101},'' {\em The Astrophysical
  Journal}~{\bf 782}(June),  L24 (2014).

\bibitem{gilhulyStellarHalosDragonfly2022a}
Gilhuly, C., Merritt, A., Abraham, R., Danieli, S., Lokhorst, D., Liu, Q., van
  Dokkum, P., Conroy, C., and Greco, J., ``Stellar {Halos} from the {The}
  {Dragonfly} {Edge}-on {Galaxies} {Survey},'' {\em ApJ}~{\bf 932},  44 (June
  2022).

\bibitem{vandokkumFORTYSEVENMILKYWAYSIZED2015}
van Dokkum, P.~G., Abraham, R., Merritt, A., Zhang, J., Geha, M., and Conroy,
  C., ``{FORTY}-{SEVEN} {MILKY} {WAY}-{SIZED}, {EXTREMELY} {DIFFUSE} {GALAXIES}
  {IN} {THE} {COMA} {CLUSTER},'' {\em ApJ}~{\bf 798},  L45 (Jan. 2015).

\bibitem{kodaApproximatelyThousandUltradiffuse2015}
Koda, J., Yagi, M., Yamanoi, H., and Komiyama, Y., ``Approximately a {Thousand}
  {Ultra}-diffuse {Galaxies} in the {Coma} {Cluster},'' {\em
  {\textbackslash}apjl}~{\bf 807},  L2 (July 2015).
\newblock \_eprint: 1506.01712.

\bibitem{janssensUltradiffuseUltracompactGalaxies2017a}
Janssens, S., Abraham, R., Brodie, J., Forbes, D., Romanowsky, A.~J., and van
  Dokkum, P., ``Ultra-diffuse and {Ultra}-compact {Galaxies} in the {Frontier}
  {Fields} {Cluster} {Abell} 2744,'' {\em ApJ}~{\bf 839},  L17 (Apr. 2017).

\bibitem{iodiceFirstDetectionUltradiffuse2020}
Iodice, E., Cantiello, M., Hilker, M., Rejkuba, M., Arnaboldi, M., Spavone, M.,
  Greggio, L., Forbes, D.~A., D'Ago, G., Mieske, S., Spiniello, C., La~Marca,
  A., Rampazzo, R., Paolillo, M., Capaccioli, M., and Schipani, P., ``The first
  detection of ultra-diffuse galaxies in the {Hydra} {I} cluster from the
  {VEGAS} survey,'' {\em Astronomy \& Astrophysics}~{\bf 642},  A48 (Oct.
  2020).
\newblock \_eprint: 2007.11533.

\bibitem{marleauUltraDiffuseGalaxies2021}
Marleau, F.~R., Habas, R., Poulain, M., Duc, P.-A., Mueller, O., Lim, S.,
  Durrell, P.~R., Sanchez-Janssen, R., Paudel, S., Ahad, S.~L., Chougule, A.,
  Bilek, M., and Fensch, J., ``Ultra diffuse galaxies in the {MATLAS}
  low-to-moderate density fields,'' {\em arXiv:2109.13173 [astro-ph]}  (Sept.
  2021).
\newblock arXiv: 2109.13173.

\bibitem{forbesUltradiffuseGalaxiesIC2020}
Forbes, D.~A., Dullo, B.~T., Gannon, J., Couch, W.~J., Iodice, E., Spavone, M.,
  Cantiello, M., and Schipani, P., ``Ultradiffuse galaxies in the {IC} 1459
  group from the {VEGAS} survey,'' {\em {\textbackslash}mnras}~{\bf 494},
  5293--5297 (June 2020).
\newblock \_eprint: 2004.10855.

\bibitem{barbosaOneHundredSMUDGes2020}
Barbosa, C.~E., Zaritsky, D., Donnerstein, R., Zhang, H., Dey, A., de~Oliveira,
  C.~M., Sampedro, L., Molino, A., Costa-Duarte, M.~V., Coelho, P., Cortesi,
  A., Herpich, F.~R., Hernandez-Jimenez, J.~A., Santos-Silva, T., Pereira, E.,
  Werle, A., Overzier, R.~A., Fernandes, R.~C., Castelli, A. V.~S., Ribeiro,
  T., Schoenell, W., and Kanaan, A., ``One hundred {SMUDGes} in {S}-{PLUS}:
  ultra-diffuse galaxies flourish in the field,'' {\em ApJS}~{\bf 247},  46
  (Mar. 2020).
\newblock arXiv: 2002.05171.

\bibitem{zaritskySystematicallyMeasuringUltraDiffuse2022}
Zaritsky, D., Donnerstein, R., Karunakaran, A., Barbosa, C.~E., Dey, A.,
  Kadowaki, J., Spekkens, K., and Zhang, H., ``Systematically {Measuring}
  {Ultra}-{Diffuse} {Galaxies} ({SMUDGes}). {III}. {The} {Southern} {SMUDGes}
  {Catalog},'' {\em arXiv e-prints} ,  arXiv:2205.02193 (May 2022).
\newblock \_eprint: 2205.02193.

\bibitem{vandokkumSpatiallyResolvedStellar2019}
van Dokkum, P., Wasserman, A., Danieli, S., Abraham, R., Brodie, J., Conroy,
  C., Forbes, D.~A., Martin, C., Matuszewski, M., Romanowsky, A.~J., and
  Villaume, A., ``Spatially {Resolved} {Stellar} {Kinematics} of the
  {Ultra}-diffuse {Galaxy} {Dragonfly} 44. {I}. {Observations}, {Kinematics},
  and {Cold} {Dark} {Matter} {Halo} {Fits},'' {\em ApJ}~{\bf 880},  91 (July
  2019).

\bibitem{vandokkumGalaxyLackingDark2018}
van Dokkum, P., Danieli, S., Cohen, Y., Merritt, A., Romanowsky, A.~J.,
  Abraham, R., Brodie, J., Conroy, C., Lokhorst, D., Mowla, L., O’Sullivan,
  E., and Zhang, J., ``A galaxy lacking dark matter,'' {\em Nature}~{\bf 555},
  629--632 (Mar. 2018).

\bibitem{danieliStillMissingDark2019}
Danieli, S., van Dokkum, P., Conroy, C., Abraham, R., and Romanowsky, A.~J.,
  ``Still {Missing} {Dark} {Matter}: {KCWI} {High}-resolution {Stellar}
  {Kinematics} of {NGC1052}-{DF2},'' {\em ApJ}~{\bf 874},  L12 (Mar. 2019).

\bibitem{lokhorstWidefieldUltranarrowbandpassImaging2020}
Lokhorst, D., Abraham, R., van Dokkum, P., and Chen, S., ``Wide-field
  ultra-narrow-bandpass imaging with the {Dragonfly} {Telephoto} {Array},'' in
  [{\em Ground-based and {Airborne} {Telescopes}
  {VIII}}{\nolinebreak\hspace{0.1em}]},  Marshall, H.~K., Spyromilio, J., and
  Usuda, T., eds.,  282, SPIE, Online Only, United States (Dec. 2020).

\bibitem{lokhorstGiantShellIonized2022}
Lokhorst, D., Abraham, R., Pasha, I., van Dokkum, P., Chen, S., Miller, T.,
  Danieli, S., Greco, J., Zhang, J., Merritt, A., and Conroy, C., ``A {Giant}
  {Shell} of {Ionized} {Gas} {Discovered} near {M82} with the {Dragonfly}
  {Spectral} {Line} {Mapper} {Pathfinder},'' {\em ApJ}~{\bf 927},  136 (Mar.
  2022).

\bibitem{racineTelescopicPointSpreadFunction1996}
Racine, R., ``The {Telescopic} {Point}-{Spread} {Function},'' {\em Publications
  of the Astronomical Society of the Pacific}~{\bf 108},  699--705 (1996).

\bibitem{bernsteinOpticalExtragalacticBackground2007}
Bernstein, R.~A., ``The {Optical} {Extragalactic} {Background} {Light}:
  {Revisions} and {Further} {Comments},'' {\em The Astrophysical Journal}~{\bf
  666}(2),  663--673 (2007).

\bibitem{devoreRetrievingCirrusMicrophysical2013}
DeVore, J.~G., Kristl, J.~A., and Rappaport, S.~A., ``Retrieving cirrus
  microphysical properties from stellar aureoles,'' {\em Journal of Geophysical
  Research: Atmospheres}~{\bf 118}(11),  5679--5697 (2013).

\bibitem{sandinInfluenceDiffuseScattered2014}
Sandin, C., ``The influence of diffuse scattered light {I}. {The} {PSF} and its
  role to observations of the edge-on galaxy {NGC} 5907,'' {\em arXiv.org} ,
  5508 (2014).

\bibitem{trujillo31MagArcsec2016a}
Trujillo, I. and Fliri, J., ``Beyond 31 mag/arcsec{\textasciicircum}2: the low
  surface brightness frontier with the largest optical telescopes,'' {\em
  ApJ}~{\bf 823},  123 (May 2016).
\newblock arXiv: 1510.04696.

\end{thebibliography}
\bibliographystyle{spiebib} 

\end{document}